\documentclass[11pt]{article}
\usepackage{fleqn,cospar}

\usepackage{url}
\def\ppn{\rp{2+2\ga-\b}{3}}

\def\eqi{\begin{equation}}
\def\eqf{\end{equation}}
\def\eqia{\begin{eqnarray}}
\def\eqfa{\end{eqnarray}}
\def\btab{\begin{tabular}}
\def\etab{\end{tabular}}
\def\bar{\begin{array}}
\def\ear{\end{array}}

\def\grl{general relativistic}

\def\leti{Lense--Thirring}

\def\se{systematic error}

\def\zh{even zonal harmonics}

\def\bm#1{{\mbox{\boldmath$#1$\unboldmath}}}
\def\gp{geopotential}

\def\lg{{\rm LAGEOS}}
\def\lgg{{\rm LAGEOS} II}

\def\lb#1{\label{#1}}
\def\pc{precession}
\def\nd{node}
\def\pg{perigee}
\def\nl{nodal}

\def\ec{eccentricity}

\def\et{Earth}
\def\ef{effect}
\def\dt#1{\dot{#1}}
\def\mlt{{\rm \mu_{LT}}}

\def\st{satellite}
\def\lt{_{\rm{LT}}}
\def\rp#1#2{{#1\over#2}}

\def\grl{general relativistic}
\def\rp#1#2{{#1\over#2}}
\def\derp#1#2{\rp{\partial{#1}}{\partial{#2}}}
\def\lb#1{\label{#1}}

\def\btab{\begin{tabular}}
\def\etab{\end{tabular}}

\def\l1{LAGEOS}
\def\l2{LAGEOS II}
\def\lt{Lense-Thirring}

\def\rfr#1{Eq.(\ref{#1})}

\def\bar{\begin{array}}
\def\ear{\end{array}}
\def\eqi{\begin{equation}}
\def\eqf{\end{equation}}

\def\jw2{w_{.2}}

\def\jn2{\dot\Omega_{.2}}

\def\jn#1{\dot\Omega_{.#1}}
\def\eqiaz{\begin{eqnarray*}}
\def\eqfaz{\end{eqnarray*}}
\def\eqia{\begin{eqnarray}}
\def\eqfa{\end{eqnarray}}
\def\btab{\begin{tabular}}
\def\etab{\end{tabular}}
\def\bar{\begin{array}}
\def\ear{\end{array}}

\def\b{\beta}
\def\ga{\gamma}

\def\et{\eta}

\def\l{\lambda}

\def\n{\nu}

\def\og{\omega}

\usepackage{graphicx}

\title{TESTING GENERAL RELATIVITY WITH SATELLITE LASER RANGING: RECENT DEVELOPMENTS}

\author{L. Iorio \address{Dipartimento di Fisica dell'Universit${\grave{a}}$ di Bari, Via Amendola 173, 70126, Bari, Italy}}

\begin{document}

\maketitle

\begin{abstract}
In this paper the most recent developments in testing General
Relativity in the gravitational field of Earth with the technique
of Satellite Laser Ranging are presented. In particular, we
concentrate our attention on some gravitoelectric and
gravitomagnetic post--Newtonian orbital effects on the motion of a
test body in the external field of a central mass. Following the
approach of the combined residuals of the orbital elements of some
existing or proposed LAGEOS--type satellites, it would be possible
to measure the gravitoelectric perigee advance with a relative
accuracy of the order of $10^{-3}$ and improve the accuracy of the
LARES mission, aimed to the measurement of the gravitomagnetic
Lense--Thirring effect, to $1\%$ or, perhaps, better. Moreover,
the use of an entirely new pair of twin LAGEOS--type satellites
placed into identical eccentric orbits with critical supplementary
inclinations would allow to measure the Lense--Thirring shift not
only by means of the sum of the nodes, with an accuracy of the
order of less than 1$\%$, but also, in a complementary way, by
means of the difference of the perigees with an accuracy of the
order of 5$\%$.
\end{abstract}

\section*{INTRODUCTION}
In the framework of the weak--field and slow--motion linearized
approximation of General Relativity (GR), we will deal with some
post--Newtonian effects of order $\mathcal{O}(c^{-2})$ on the
orbits of test bodies freely falling in the terrestrial
gravitational field and will examine recent developments about the
possibility of measuring them by means of the Satellite Laser
Ranging (SLR) technique.
\subsection*{The Gravitoelectric Effects}
The geodetic, or De Sitter, precession (De Sitter, 1916) refers to
the coupling of the static, gravitoelectric part of the
gravitational field due to the Schwarzschild metric generated by a
central, non--rotating mass to the spin of a particle freely
orbiting around it. It has been measured for Earth--Moon orbit,
thought of as a giant gyroscope, in the gravitational field of Sun
with 1$\%$ accuracy (Williams et al., 1996). It should be measured
for four superconducting gyroscopes in the gravitational field of
Earth by the important GP-B mission (Everitt et al., 2001) at a
claimed relative accuracy level of $2\times 10^{-5}$. However,
such mission has recently experienced some problems (Lawler,
2003a; 2003b).

Perhaps the most famous of the gravitoelectric post--Newtonian
orbital effects, the advance of the pericentre $\omega$ of the
orbit of a test particle (i.e., non--rotating) due to the
Schwarzschild part of the metric of a central, non--rotating mass
(Ciufolini and Wheeler, 1995) has represented one of the classical
tests of GR thanks to the measurement of the perihelion shifts of
Mercury and other planets in the gravitational field of Sun with
the radar ranging technique. The relative accuracy of such
measurements is of the order of 10$^{-2}$--10$^{-3}$ (Shapiro,
1990). Here we will present a proposal for obtaining a
complementary measurement of such orbital effect in the
gravitational field of Earth by using some existing or proposed
SLR satellites of LAGEOS--type. The achievable relative accuracy
should amount to 10$^{-3}$ (Iorio et al., 2002a), according to the
present--day knowledge of the models of the gravitational and
non--gravitational forces acting on the satellites of
LAGEOS--type.
\subsection*{The Gravitomagnetic Effects}
Contrary to the classical Newtonian mechanics, GR predicts that
the gravitational field of a central mass is sensitive to its
state of motion. In particular, if it rotates and has a proper
angular momentum \bm J, it turns out that a freely falling body
orbiting it is affected by many so called gravitomagnetic effects.
Such name is due to the close formal analogies of the linearized
approximation of GR with the Maxwellian electromagnetism. For a
recent review see (Ruggiero and Tartaglia, 2002). In regard to
such gravitomagnetic effects, they could be divided into two main
categories: those induced by the spin--spin coupling and those
induced by the spin--orbit coupling.
\subsubsection*{The Gravitomagnetic Spin--Spin Effects}
When a spinning particle moves in an external gravitational field
one has to describe both its spin precession and the influence of
the spin on its trajectory (Khriplovich, 2001).

When the spin {\bm J} of the central object is taken into account,
it turns out that it affects the precessional motion of the spin
{\bm s} of a particle freely orbiting it in a way discovered by
Schiff (Schiff, 1960) in 1960. The detection of this subtle
precessional effect, in addition to the gravitoelectric geodetic
precession of the spin, is one of the most important goals of the
GP-B mission (Everitt et al., 2001); the claimed accuracy amounts
to $1\%$.

In regard to possible orbital effects on the motion of a spinning
particle freely orbiting a central rotating mass in an
astrophysical context, they have been recently calculated in
(Iorio, 2003a). Unfortunately, they turn out to be too small to be
detected in Solar System space--based experiments. See also (Zhang
et al., 2001) for a different, phenomenological approach not a
priori based on GR, which has been applied to some preliminary
Earth--based laboratory tests (Zhou et al., 2002).
\subsubsection*{The Gravitomagnetic Spin--Orbit Effects}
The Lense--Thirring drag of the inertial frames (Lense and
Thirring, 1918) is an effect due to the stationary,
gravitomagnetic part of the gravitational field of a central
rotating mass on the geodesic path of a test particle. Such effect
could be thought of as a spin--orbit interaction between the spin
of the central object and the orbital angular momentum \bm {\ell}
of the test body. For some spin--orbit effects induced by the
rotation of Sun on the orbital angular momentum of Earth--Moon
system see (Mashhoon and Theiss, 2001). In 1998 the first evidence
of the Lense--Thirring effect in the gravitational field of Earth
has been reported (Ciufolini et al., 1998) with a claimed accuracy
of almost $20\%$. It is based on the analysis of a suitable
combination of the orbital residuals of the nodes $\Omega$ of the
LAGEOS and LAGEOS II SLR satellites and of the perigee $\omega$ of
LAGEOS II.

Another interesting gravitomagnetic effect of order
$\mathcal{O}(c^{-2})$ on the orbit of a test particle has been
recently derived in (Iorio, 2002a); it is induced by the temporal
variability of the angular momentum of the central mass.
Unfortunately, it is too small to be detected with SLR in the
terrestrial gravitational field.

In this paper we will show how the launch of the proposed
LAGEOS--like LARES satellite could allow to measure the
Lense--Thirring effect with an accuracy probably better than $1\%$
(Iorio et al., 2002b). Moreover, the concept of twin SLR
satellites placed into identical orbits in orbital planes with
supplementary inclinations will be extended to new observables and
a possible mission involving the use of an entirely new pair of
SLR satellites of LAGEOS--type will be sketched.
\section{THE MAJOR SYSTEMATIC ERRORS}
In all the performed or proposed experiments with which we will
deal in this paper the reliable assessment of the error budget is
of the utmost importance. Indeed, the terrestrial space
environment is rich of competing classical perturbing forces of
gravitational and non--gravitational origin which in many cases
are far larger than the general relativistic effects to be
investigated. In particular, it is the impact of the systematic
errors induced by the mismodelling in such various classical
perturbations which is relevant in determining the total realistic
accuracy of an experiment like those previously mentioned.

The general relativistic effects of interest here are linear
trends affecting the node $\Omega$ and/or the perigee $\omega$ of
the orbit of a satellite. A LAGEOS--type satellite's orbit is
affected by them at a level of $10^1$--$10^3$ milliarcseconds per
year (mas/y in the following) for the gravitomagnetic and the
gravitoelectric effects, respectively.

In this context the most important source of systematic error is
represented by the classical secular precessions of the node and
the perigee induced by the mismodelled even ($l=2n,\ n=1,2,3,...$)
zonal ($m=0$) harmonic coefficients $J_2,\ J_4,\ J_6,...$ of the
multipolar expansion of the terrestrial gravitational field,
called geopotential. Indeed, while the time--varying tidal orbital
perturbations ({Iorio}, 2001; { Iorio and Pavlis}, 2001; { Pavlis
and Iorio}, 2002) and non--gravitational orbital perturbations
(Lucchesi, 2001; 2002), according to their periods $P$ and to the
adopted observational time span $T_{\rm obs}$, can be viewed as
empirically fitted quantity and can be removed from the signal,
this is not the case of the classical even zonal secular
precessions. Their mismodelled linear trends act as superimposed
effects which may alias the recovery of the genuine general
relativistic features. Such disturbing trends cannot be removed
from the signal without cancelling also the general relativistic
signature, so that one can only assess as more accurately as
possible their impact on the measurement. The systematic error
induced by the mismodelled part of the geopotential can then be
viewed as a sort of unavoidable part of the total systematic
error.

The same considerations hold also for the aliasing secular trends
induced by some tiny non--gravitational thermal perturbations like
the terrestrial Yarkovsky--Rubincam effect (Lucchesi, 2002).
\subsection{The Systematic Error due to the Geopotential} A possible strategy for
reducing the impact of the error due to the geopotential, as we
will see in the following sections, consists of suitable
combinations of the orbital residuals $\delta\dot\Omega$ and
$\delta\dot\omega$ of the rates of the nodes and the perigees of
different SLR satellites. Such combinations can be written in the
form \eqi\sum_{i=1}^N c_i f_i=X_{\rm GR}\mu_{\rm
GR},\lb{combi}\eqf in which the coefficients $c_i$ are, in
general, suitably built up with the orbital parameters of the
satellites entering the combinations, the $f_i$ are the residuals
$\delta\dot\Omega,\ \delta\dot\omega$ of the rates of the nodes
and the perigees of the satellites entering the combination,
$X_{\rm GR}$ is the slope, in mas/y, of the general relativistic
trend of interest and $\mu_{\rm GR}$ is the solve--for parameter,
to be determined by means of usual least--square procedures, which
accounts for the general relativistic effect. For example, in the
case of the Lense--Thirring--LAGEOS--LAGEOS II experiment
(Ciufolini, 1996) $X_{\rm LT}=60.2$ mas/y, while for the
gravitoelectric perigee advance (Iorio et al., 2002a) $X_{\rm
GE}=3,348$ mas/y. More precisely, the combinations of \rfr{combi}
are obtained in the following way. The equations for the residuals
of the rates of the $N$ chosen orbital elements are written down,
so to obtain a non homogeneous algebraic linear system of $N$
equations in $N$ unknowns. They are $\mu_{\rm GR}$ and the first
$N-1$ mismodelled spherical harmonics coefficients $\delta J_l$ in
terms of which the residual rates are expressed. The coefficients
$c_i$ and, consequently, $X_{\rm GR}$ are obtained by solving for
$\mu_{\rm GR}$ the system of equations. So, the coefficients $c_i$
are calculated in order to cancel out the contributions of the
first $N-1$ even zonal mismodelled harmonics which represent the
major source of uncertainty in the Lense--Thirring and
gravitoelectric precessions ({Ciufolini}, 1996; Iorio, 2002b;
Iorio et al., 2002a; 2002b). The coefficients $c_i$ can be either
constant \footnote{In general, the coefficient of the first
orbital element entering a given combination is equal to 1, as for
the combinations in ({Ciufolini}, 1996; {Iorio}, 2002b; Iorio et
al., 2002a; 2002b). } or depend on the orbital elements of the
satellites entering the combinations.

Now we expose how to calculate the systematic error due to the
mismodelled even zonal harmonics of the geopotential for the
combinations involving the residuals of the nodes and the perigees
of various satellites.

In general, if we have an observable $q$ which is a function
$q=q(x_j)$, $j=1,2...M$ of $M$ correlated parameters $x_j$ the
error in it is given by
 \eqi \delta
q=\left[\sum_{j=1}^M\left(\derp{q}{x_j}\right)^{2}\sigma_j^2+2\sum_{h\neq
k
=1}^M\left(\derp{q}{x_h}\right)\left(\derp{q}{x_k}\right)\sigma^{2}_{hk}\right]^{\frac{1}{2}}\lb{app1}\eqf
in which $\sigma^{2}_{j}\equiv C_{jj}$ and $\sigma^{2}_{hk}\equiv
C_{hk}$ where $\{C_{hk}\}$ is the square matrix of covariance of
the parameters $x_j$.

In our case the observable $q$ is any residuals' combination \eqi
q=\sum_{i=1}^{N}c_i f_i(x_j),\ j=1,2...10,\eqf where $x_j,\
j=1,2...10$ are the even zonal geopotential's coefficients $J_2,\
J_4...J_{20}$.  Since \eqi \derp{q}{x_j}=\sum_{i=1}^N c_i
\derp{f_i}{x_j},\ j=1,2...10\lb{app2},\eqf by putting \rfr{app2}
in \rfr{app1} one obtains, in mas/y \eqi \delta
q=\left[\sum_{j=1}^{10}\left(\sum_{i=1}^N c_i
\derp{f_i}{x_j}\right)^{2}\sigma_j^2+2\sum_{h\neq k
=1}^{10}\left(\sum_{i=1}^N c_i
\derp{f_i}{x_h}\right)\left(\sum_{i=1}^N c_i
\derp{f_i}{x_k}\right)\sigma^{2}_{hk}\right]^{\frac{1}{2}}.\lb{app3}\eqf
The percent error, for a given \grl\ trend and for a given
combination, is obtained by taking the ratio of \rfr{app3} to the
slope in mas/y of the \grl\ trend for the residual combination
considered.

The validity of \rfr{app3} has been checked by calculating with it
and the covariance matrix of the EGM96 gravity model (Lemoine et
al., 1998) up to degree $l=20$ the systematic error due to the
even zonal harmonics of the geopotential of the Lense--Thirring
LAGEOS--LAGEOS II experiment; indeed the result \eqi\delta\mu_{\rm
LT}=13\%\ \mu_{\rm LT}\eqf claimed in ({Ciufolini et al.}, 1998)
has been obtained again. For the systematic errors due to the even
zonal harmonics of the geopotential of alternative proposed
gravitomagnetic and gravitoelectric experiments, according to
EGM96, see ({Iorio}, 2002b; {Iorio et al.}, 2002a; 2002b). It is
worth noticing that, since the orbits of the LAGEOS satellites are
almost insensitive to the geopotential's terms of degree higher
than $l=20$, the estimates based on the covariance matrix of EGM96
up to degree $l=20$ should be considered rather reliable although
the higher degree terms of EGM96 might be determined with a low
accuracy.

It should also be pointed out that the evaluations of the
systematic error due to geopotential based on the approach of the
combined orbital residuals should be free from some uncertainties
due to possible seasonal effects. Indeed, according to (Ries et
al., 2003), the covariance matrix of EGM96 (and of other previous
gravity models as well) would not yield reliable results for a
generic time span of a few years, but only just for the temporal
interval during which the data used for its construction have been
collected or, at most, for some very long, averaged time span. It
would be so because of the impact of possible secular, seasonal
and stochastic variations of the terrestrial gravity field which
would have not been accounted for in the model solution. However,
since it turns out that such seasonal effects would mainly affect
just the first even zonal harmonic coefficients of the
geopotential, the uncertainty related to them should be very small
for residual combinations which, by construction, cancel out just
the first even zonal harmonic coefficients of the geopotential. On
the other hand, if we cancel out as many even zonal harmonics as
possible, the uncertainties in the evaluation of the systematic
error based on the remaining correlated even zonal harmonics of
higher degree should be greatly reduced, irrespectively of the
chosen time span. Moreover, also certain time--dependent harmonic
perturbations of gravitational origin are canceled out. I.e., the
most insidious of such perturbations is the 18.6--year tide which
is just a $l=2,\ m=0$ constituent with a period of 18.6 years due
to the motion of the Moon's node and which affects both the nodes
and the perigees of the LAGEOS satellites at a level of 10$^3$ mas
(Iorio, 2001). It is cancelled out by the combinations of orbital
residuals.

A very important point to stress is that the forthcoming new data
on Earth gravitational field by CHAMP (Pavlis, 2000), which has
been launched in July 2000, and especially GRACE (Ries et al.,
2002), which has been launched in March 2002, should have a great
impact on the reduction of the systematic error due to the
mismodelled part of geopotential, provided that the great
expectations about the level of achievable improvement of our
knowledge of the terrestrial gravity field will be finally
satisfied. In particular, if the other existing LAGEOS--type
satellites have to be included in the combined residual
combinations, it would be very important that the new terrestrial
gravity models improve the accuracy in determining the even zonal
harmonics of higher degree to which they are sensitive, contrary
to the LAGEOS satellites. The first, very preliminary Earth
gravity models including some data from CHAMP and GRACE seem to
confirm such expectations (Iorio and Morea, 2003).
\section*{MEASURING THE GRAVITOELECTRIC PERIGEE ADVANCE WITH SLR}
The gravitoelectric secular rate of the pericentre of a test body
freely orbiting a central static mass is (Ciufolini and Wheeler,
1995) \eqi \dot\og_{\rm GE}=\rp{3 n G M}{c^2 a
(1-e^2)}\times\ppn,\lb{peg}\eqf in which $G$ is the Newtonian
gravitational constant, $c$ is the speed of light in vacuum, $M$
is the mass of the central object, $a$ and $e$ are semimajor axis
and eccentricity, respectively, of the orbit of the test body and
$n=\sqrt{GM/a^{\rm 3}}$ is its mean motion. In the following we
define $\n\equiv \ppn$, where $\gamma$ and $\beta$ are the
Eddington--Robertson--Schiff PPN parameters (Will, 1993) which
test the alternative metric theories of gravitation. The
gravitoelectric precessions for the \lg\ satellites, according to
GR ($\beta=\gamma=1$) amount to \eqi \dot\omega_{\rm GE}^{\rm
LAGEOS} =  3,312.4\ \textrm{mas/y},\ \dot\omega_{\rm GE}^{\rm
LAGEOS\ II}  =  3,348.5\ \textrm{mas/y}. \eqf In regard to the
possibility of measuring such rates in the terrestrial
gravitational field with SLR, in (Ciufolini and Matzner, 1992) a
first attempt using the perigee of LAGEOS only is reported, but
the estimated total systematic error is of the order of 20$\%$.
\subsection*{The Use of LAGEOS and LAGEOS II}
If the perigee of LAGEOS or LAGEOS II only was used, the
systematic relative error due to the geopotential, according to
the covariance matrix of the even zonal harmonics up to the degree
$l=20$ of EGM96, would amount to $8\times 10^{-3}$ and $2\times
10^{-2}$, respectively. Instead, following the approach of the
combined residuals outlined in the previous section it is possible
to adopt as observable (Iorio 2002b; Iorio et al., 2002a)
\eqi\delta\dot\omega^{\rm LAGEOS\ II}+k_1\delta\dot\Omega^{\rm
LAGEOS\ II}+k_2\delta\dot\Omega^{\rm LAGEOS} =3,348.46\times\n
,\lb{rsd}\eqf with \eqi k_1  =  -0.87,\ k_2  = -2.86,\lb{cc2}\eqf
so that we could exploit the insight acquired with the \lt\
\lg--LAGEOS II experiment. The systematic relative error due to
the geopotential, according to the covariance matrix of the even
zonal harmonics of EGM96 up to the degree $l=20$, amounts to
$6\times 10^{-3}$. The total systematic relative error over a time
span of 8 years, including other sources of error, amounts to
almost 7$\times 10^{-3}$ (Iorio et al., 2002a). As a secondary
outcome of such experiment, it would be possible to use its
results with those for $\et=4\beta-\gamma -3$ from Lunar Laser
Ranging (LLR) (Anderson and Williams, 2001) in order to get
independent measurements of $\beta$ and $\gamma$ with an accuracy
of $3\times 10^{\rm -3}$ and $1\times 10^{\rm -2}$, respectively.
The estimates for $\nu$, $\beta$ and $\gamma$ should be improved
by the new forthcoming gravity models from the CHAMP and GRACE
missions.
\section*{A REVISITED VERSION OF THE LARES MISSION}
The Lense--Thirring secular rates of the node and the perigee of a
test body freely orbiting a central rotating mass are ({Lense and
Thirring}, 1918) \eqi \dot\Omega _{\rm LT}  =
\frac{2GJ}{c^{2}a^{3}(1-e^{2})^{\frac{3}{2}}},\ \dot\omega_{\rm
LT}  =
-\frac{6GJ\cos{i}}{c^{2}a^{3}(1-e^{2})^{\frac{3}{2}}},\lb{letiperig}\eqf
where $J$ is the proper angular momentum of the central mass and
$i$ is the inclination of the orbit of the test particle to the
equator of $M$. The
\leti\ precessions for the \lg\ satellites amount to \eqi
\dot\Omega_{\rm LT}^{\rm LAGEOS} =  31\ \textrm{mas/y},\
\dot\Omega_{\rm LT}^{\rm LAGEOS\ II}  =  31.5\ \textrm{mas/y},\
\dot\omega_{\rm LT}^{\rm LAGEOS}  =  31.6\ \textrm{mas/y},\
\dot\omega_{\rm LT}^{\rm LAGEOS\ II}  =  -57\ \textrm{mas/y}. \eqf

The first measurement of this \ef\ in the gravitational field of
Earth has been obtained by analyzing a suitable combination of the
laser-ranged data to the existing SLR \lg\ and \lgg\ satellites
({Ciufolini et al.,} 1998). The observable is a linear trend with
a slope of 60.2 mas/y and includes the residuals of the nodes of
\lg\ and \lgg\ and the \pg\footnote{The \pg\ of \lg\ was not used
because it introduces large observational errors due to the
smallness of the \lg\ \ec\ (Ciufolini, 1996) which amounts to
0.0045.} \ of \lgg (Ciufolini, 1996). The claimed total relative
accuracy of the measurement of the solve-for parameter $\mlt$,
introduced in order to account for this \grl\ \ef, is about
$2\times 10^{-1}$ ({Ciufolini et al.}, 1998).

As already pointed out, in this kind of experiments using Earth
satellites the major source of \se s is represented by the
aliasing trends due to the uncancelled classical secular
precessions of the \nd\ and the \pg\ induced by the mismodelled
\zh\ of the \gp\ $J_2,\ J_4,\ J_6,...$.

In order to achieve a few percent accuracy, in ({Ciufolini}, 1986)
it was proposed to launch a passive geodetic laser-ranged \st- the
former {\rm LAGEOS} III - with the same orbital parameters of \lg\
($a=12,270$ km, $e=0.0045$, $i=110^{\circ}$) apart from its
inclination which should be supplementary to that of \lg, i.e.
$i_{\rm LAGEOS\ III}=70^{\circ}$.

This orbital configuration would be able to cancel out exactly the
classical \nl\ \pc s, which are proportional to $\cos i$ (Iorio,
2002c), provided that the observable to be adopted is the sum of
the residuals of the \nl\ \pc s of {\rm LAGEOS} III and LAGEOS
\eqi \delta\dt\Omega^{{\rm LAGEOS\ III}}+\delta\dt\Omega^{{\rm
LAGEOS}}=62\mlt.\lb{lares}\eqf Later on the concept of the mission
slightly changed. The area-to-mass ratio of {\rm LAGEOS} III was
reduced in order to make less relevant the impact of the
non-gravitational perturbations, the total weight of the satellite
was reduced to about 100 kg, i.e. to about 25$\%$ of the weight of
LAGEOS, and the eccentricity was enhanced to $e_{\rm LR}=0.04$ in
order to be able to perform other \grl\ tests: the LARES was born
({Ciufolini}, 1998). At present the LARES experiment has not yet
been approved by any space agency. Although much cheaper than
other proposed and approved complex space--based missions, funding
is the major obstacle in implementing the LARES project.
\subsection*{Some Possible Weak Points of the Originally Proposed Lares Mission}
Since the eccentricities of LAGEOS and LARES are 0.0045 and 0.04,
respectively, the cancellation of the classical secular nodal
precession amounts to 0.3$\%$, according to \rfr{lares} applied to
to the LAGEOS--LARES configuration and the covariance matrix of
the even zonal harmonics of EGM96 up to the degree $l=20$ and
using the nominal values of the orbital parameters of LARES.

The major drawback of \rfr{lares}, applied to the LAGEOS--LARES
configuration, is that it turns out to be rather sensitive to the
possible departures of the LARES orbital parameters from their
nominal values due to the unavoidable orbital injection errors.
More precisely, for deviations of just 1$^{\circ}$ from the
projected nominal supplementary configuration with LAGEOS, the
systematic error due to the even zonal harmonics of the
geopotential would amount to 1-2$\%$. Of course, this fact would
put rather stringent constraints on the quality and, consequently,
the cost of the LARES launcher.
\subsection*{A New, More Accurate Residuals Combination}
Following the strategy of the combined orbital residuals
previously outlined, it could be possible to adopt the following
observable \eqi \delta\dt\Omega^{\rm {\rm
LAGEOS}}+c_1\delta\dt\Omega^{\rm {\rm LAGEOS}\
II}+c_2\delta\dt\Omega^{\rm LARES}+c_3\delta\dt\omega^{\rm {\rm
LAGEOS} \ II}+c_4\delta\dt\omega^{\rm
LARES}=61.8\mlt,\lb{combils}\eqf with \eqi c_1  =  6\times
10^{-3}, \ c_2  =  9.83\times 10^{-1},\ c_3  =  -1\times 10^{-3},\
c_4  =  -2\times 10^{-3}\lb{c4}. \eqf According to the covariance
matrix of the even zonal harmonics of EGM96 up to the degree
$l=20$, the systematic error in \rfr{combils} due to the
geopotential reduces to 0.02$\%$, i.e one order of magnitude
better than \rfr{lares}. Moreover, and this is a crucial point, it
turns out that such result is completely insensitive to the
possible departures of the LARES orbital parameters from their
nominal values, especially in regard to the inclination $i_{\rm
LR}$, contrary to \rfr{lares}. This would allow to use a rather
cheap launcher for LARES. In (Iorio, 2003d) it has been shown also
that the correlations among the various even zonal harmonics are
not relevant for the error in \rfr{combils}.

In regard to the time--dependent gravitational and
non--gravitational perturbations which sensibly affect the
perigee, it should be noticed that the perigees of LAGEOS II and
LARES are weighted, in \rfr{combils}, by coefficients of order of
$10^{-3}$, so that the impact of the time--varying harmonic
perturbations on $\omega$ would be notably reduced\footnote{This
result is really important, especially in view of possible new
phenomena in the surface properties of the LAGEOS satellites which
might affect the perigee, as it seems to happen for LAGEOS II
(Iorio et al., 2002b). }.

According to the estimates of (Iorio et al., 2002b), the total
systematic error should be $\leq 1\%$. Of course, when the new,
hopefully more accurate data on the static and dynamical parts of
the terrestrial gravitational field from the CHAMP and GRACE
missions will be available, a substantial improvement in such
estimate should occur.
\subsection*{Using LARES for the Measurement of the Gravitoelectric Perigee Advance}
It would also be possible to include the data of LARES in some
combinations of orbital residuals in order to measure the
gravitoelectric perigee advance. E.g., a possible observable is
\eqi\delta\dot\omega^{\rm LAGEOS\ II}+w_1\delta\dot\omega^{\rm
LARES}+w_2\delta\dot\Omega^{\rm LAGEOS\ II}=X_{\rm GE}\mu_{\rm GE
},\lb{larespg}\eqf with \eqi w_1  =  -4.71,\ w_2  =  2.26,\ X_{\rm
GE}  =  -12,117.4\ {\rm mas/y}. \eqf The impact of the uncancelled
even zonal harmonics, with degree higher than four, amounts to an
uncertainty of about $6\times 10^{-3}$, according to the
covariance matrix of the even zonal harmonics up to the degree
$l=20$ of EGM96. This error should be further reduced when the new
gravity models from the CHAMP and GRACE missions will be
available. An interesting feature of \rfr{larespg} is that the
impact of the non--gravitational perturbations would produce an
uncertainty of about $4\times 10^{-3}$ over a time span of 7
years, according to very conservative estimates (Iorio et al.,
2002a); in particular the Earth thermal thrust, i.e. the so-called
Yarkovski--Rubincam effect, would induce a mismodeled secular
trend with an uncertainty of only $6\times 10^{-5}$. For \rfr{rsd}
the error due to the non--gravitational perturbations amounts to
$1\times 10^{-2}$, while the impact of the Yarkovski--Rubincam
effect amounts to $1\times 10^{-4}$. The inclusion of LARES would
thus represent an improvement with respect to the LAGEOS and
LAGEOS II scenario of \rfr{rsd}, especially when, in the near
future, the impact of the mismodeled non--gravitational
perturbations will increase relatively to the gravitational
perturbations thanks to the more accurate Earth  gravitational
field models.
\section*{NEW OBSERVABLES FOR THE SUPPLEMENTARY SATELLITES CONFIGURATION}
Up to now, the concept of twin LAGEOS--type satellites in
identical orbits with supplementary inclinations has been used in
the context of the originally proposed LAGEOS--LARES mission whose
observable is the sum of the nodes of \rfr{lares}.
\subsection*{The Difference of the Perigees}
It turns out that such concept could be fruitfully extended to
another independent observable sensitive to the gravitomagnetic
Lense--Thirring effect. Indeed, for a pair of satellites in such
orbital configuration, according to (Iorio, 2002c) and to
\rfr{letiperig}, on one hand the classical secular precessions of
the perigee due to the even zonal harmonics of the geoptential are
equal because they depend on $\cos^2 i$ and on even powers of
$\sin i$, on the other the Lense--Thirring rates of the perigee
are equal and opposite because they depend on $\cos i$. So,
besides the sum of the nodes, also the difference of the perigees
\eqi \delta\dot\omega^{i }-\delta\dot\omega^{180^{\circ}-i}=X_{\rm
LT}\mu_{\rm LT}\lb{difzaperi}\eqf could be considered as another
independent, although less precise, gravitomagnetic observable
(Iorio, 2003b; 2003c).

Of course, such observable could not be used in the LARES mission
because the perigee of LAGEOS is not well defined. Moreover, it
would neither be a good idea to think about a possible LARES II
satellite supplementary to LAGEOS II because the perigee of LAGEOS
II is affected by some gravitational and non--gravitational
time--varying perturbations with periods of many years (Iorio,
2001; Lucchesi 2001), so that over reasonable time spans of only a
few years they would resemble superimposed biasing linear trends
which would corrupt the measurement of the genuine Lense--Thirring
secular trend.

It turns out that an entirely new pair of twin LAGEOS--type
satellites in identical eccentric orbits ($a=12,000$ km, $e=0.05$)
with supplementary critical inclinations ($i^{(1)}=63.4^{\circ}$,
$i^{(2)}=116.6^{\circ}$) would be a good choice. Indeed, the so
obtained frozen perigee configuration\footnote{For
$i=63.4^{\circ}$ the classical secular precession on $\omega$ due
to $J_2$ vanishes.} would allow to reduce the periods of all the
time--dependent harmonic perturbations, so that they could be
easily fitted and removed from the signal over a time span of a
few years. Moreover, since the eccentricities would be the same,
the cancellation of the mismodelled classical secular precessions
due to the even zonal harmonics of the geoptential would occur at
an higher level of accuracy than that of the originally proposed
LAGEOS--LARES mission.

According to numerical estimates based on the physical and
geometrical properties of the existing LAGEOS satellites, the
total accuracy over an observational time span of 6 years would be
of the order of $5\%$ (Iorio and Lucchesi, 2003). In it a very
important role would be played by the non--gravitational
perturbations. It is worth noticing that their importance in the
determination of the total error budget will greatly increase when
the new, more accurate data for the static and time--varying parts
of Earth gravitational field will be available from the CHAMP and,
especially, GRACE missions.

Of course, such new configuration would allow to increase the
precision of the sum of the nodes as well. According to very
preliminary numerical estimates based on the physical and
geometrical properties of the LAGEOS satellites, the total
accuracy over an observational time span of 6 years would be
$\leq$ $1\%$ (Iorio and Lucchesi, 2003).

The addition of the sum of the nodes and of the difference of the
perigees \eqi\delta\dot\Omega^{i
}+\delta\dot\Omega^{180^{\circ}-i}+\delta\dot\omega^{i
}-\delta\dot\omega^{180^{\circ}-i}=X_{\rm LT}\mu_{\rm LT} \eqf
would increase the accuracy to 2.8$\%$ (Iorio and Lucchesi, 2003).
Such estimates should further improve because they do not account
for the possibility of fitting and removing some time--dependent
orbital perturbations with known periodicities from the time
series.
\subsection*{The Combined Residuals Scenario}
If, instead of LARES, we would use the data of one of the two new
proposed satellites with critical inclination in some combinations
of orbital residuals we could measure both the gravitoelectric
perigee advance and the Lense--Thirring effect. A possible
observable would be
\eqi\delta\dot\omega^{i}+p_1\delta\dot\Omega^{\rm LAGEOS\ II
}+p_2\delta\dot\Omega^{\rm LAGEOS} +p_3\delta\dot\omega^{\rm
LAGEOS\ II}+p_4\delta\dot\Omega^{i}=\mu_{\rm GR }X_{\rm
GR},\lb{combigraz}\eqf in which \eqi p_1 = -1.55, \ p_2 = -2.77, \
p_3 = 0.348, \ p_4 = 0.361. \eqf For the gravitoelectric perigee
advance we would have $X_{\rm GE}=4,636.5$ mas/y and a systematic
relative error due to the geopotential of $2\times 10^{-4}$,
according to the covariance matrix of the even zonal harmonics of
EGM96 up to the degree $l=20$. For the gravitomagnetic
Lense--Thirring effect we would have $X_{\rm LT}=-187$ mas/y and a
systematic relative error due to the geopotential of $5\times
10^{-3}$, according to the covariance matrix of the even zonal
harmonics of EGM96 up to the degree $l=20$. In regard to the
measurement of the Lense--Thirring effect, the combination of
\rfr{combils} involving LARES, LAGEOS and LAGEOS II would be more
accurate than \rfr{combigraz}. On the other hand, in regard to the
measurement of the gravitoelectric perigee adavance,
\rfr{combigraz} would represent a notable improvement with respect
to \rfr{larespg}.
\section*{CONCLUSIONS}
In this paper we have illustrated some new recent developments in
the field of precise testing GR with SLR.

By suitably combining the nodes of LAGEOS and LAGEOS II and the
perigee of LAGEOS II it would be possible to obtain a
complementary measurement of the gravitoelectric perigee advance
in the gravitational field of Earth with the SLR technique at an
accuracy level of 0.7$\%$.

A modified version of the observable to be used in the LARES
mission, which includes not only the node of LAGEOS but also the
node and the perigee of LAGEOS II and the perigee of LARES in a
suitable combination, should be able to measure the
gravitomagnetic Lense--Thirring effect at an accuracy level of the
order of 1$\%$ by using a not too expensive launcher for LARES.
Indeed, the error due to the even zonal harmonics of the new
proposed observable would be insensitive to the unavoidable
orbital injection errors, to the correlations among the even zonal
harmonic coefficients and, to a certain extent, to the adopted
Earth gravity model.

The use of a couple of entirely new SLR LAGEOS--type satellites
placed into identical eccentric orbits with critical supplementary
inclinations would allow to use not only the sum of the nodes but
also the difference of the perigees as independent observables for
the measurement of the Lense--Thirring effect. According to
preliminary numerical estimates based on the properties of the
existing LAGEOS satellites, the accuracy would be of the order of
$1\%$ and $5\%$, respectively.

When the new, more accurate data for the terrestrial gravitational
field from the CHAMP and, especially, GRACE missions will be
available, such estimates should be notably improved. The first,
very preliminary, Earth gravity models including some data of
CHAMP and GRACE seem to confirm such expectations.


\noindent
E-mail address of L.Iorio\ \ \ \ Lorenzo.Iorio@ba.infn.it\\
Manuscript received 19 October 2002
\end{document}